\documentclass[floatfix,aps,twocolumn,prl,superscriptaddress]{revtex4-1}
\usepackage{graphicx}
\usepackage{dcolumn}
\usepackage{bm}
\usepackage[T1]{fontenc}
\usepackage[latin9]{inputenc}
\usepackage{textcomp}
\usepackage{amsmath}
\usepackage{mathrsfs}
\usepackage{graphicx}
\usepackage{amssymb}
\usepackage{units}
\usepackage{xcolor}
\usepackage{float}

\newcommand{\iac}[1]{#1}

\begin{document}

\title{Bogoliubov theory of the laser linewidth and application to polariton condensates}

\author{Ivan Amelio}
\affiliation{
Institute of Quantum Electronics ETH Zurich, CH-8093 Zurich, Switzerland}
\author{Iacopo Carusotto}
\affiliation{INO-CNR BEC Center and Dipartimento di Fisica, Universit{\`a} di Trento, 38123 Povo, Italy}

\begin{abstract}
For a generic semi-classical laser dynamics in the complex Ginzburg-Landau form,
we develop a Bogoliubov approach for the computation of the
laser emission linewidth.
Our method provides a unifying perspective of the treatments by Henry and Petermann: 
both broadening mechanisms are ascribed to  the non-orthogonality of the Bogoliubov modes, which live in a space with doubled degrees of freedom.
As an  example of application, the method allows to study  the 
interplay of driven-dissipation, interactions and spatial inhomogeneity typical of 
polariton condensates.
The traditional theory of the Henry and Petermann factors is found to fail dramatically in  the presence of sizable polariton-polariton interactions.
In particular, also in a strong confining potential, the intrinsically
 multi-mode nature of the density fluctuations  has to be considered in order to  describe quantitatively phase diffusion {\em \`a la} Henry. 
\end{abstract}
\date{\today}

\maketitle


\section{Introduction}

The emission spectrum of a lasing system reflects the complex dynamics of spatial and temporal fluctuations.
For a single mode laser,    Schawlow and Townes predicted 
the linewidth of the
Lorentzian spectral peak
to scale as the inverse of the emitted power~\cite{schawlow1958}. 
However, corrections to this simple formula have been   
first proposed by Petermann~\cite{petermann1979}, who noticed that deviations of the
shape of the 
lasing mode from 
the  eigenmode of a lossless cold cavity determine  a broadening of the laser linewidth~\cite{henry1986,siegman1989a,siegman1989b,hamel1989,
hamel1990,grangier1998}. 
Then, the crucial role of the fluctuations of the intensity, hence of the nonlinear refractive index,  was first considered by Henry close to threshold~\cite{henry1982}, but only in the last decade 
nonlinearities have been fully included in the treatment of the linewidth~\cite{chong2012,pick2015, pick2019}.
As a further broadening mechanism, we have recently shown \cite{amelio2020theory} that in low dimensions the lack of long-range order can determine a  dramatic broadening of the linewidth, which will not scale with the inverse of the number of photons in the system as predicted by Schawlow and Townes.
In particular, in one dimension the scaling is determined by universal Kardar-Parisi-Zhang exponents~\cite{ameliochiocchetta}. 
Finally, for large semiconductor arrays and random lasers it can occur that many modes  lase together and the spectra become complex~\cite{tureci2006,tureci2008}. 

In this work we develop a theoretical framework for the computation of the  linewidth of a generic  semi-classical laser dynamics via the solution of a Bogoliubov eigenproblem. The formalism allows for an efficient and transparent study of broadening effects of the Henry and Petermann type
for a laser device operating in the single mode regime.  Multimode lasing,  instead, is currently not included in our treatment and KPZ effects~\cite{amelio2020theory} are intrinsically beyond any linear response approach.
The Bogoliubov method has been originally developed for the computation of the collective modes of quantum fluids~\cite{pitaevskii2016}  and
can be applied to a remarkably broad range of physical situations, including non-equilibrium fluids of light~\cite{carusotto2013}, acoustic Hawking radiation
~\cite{recati2009}
and 
(disordered) topological lasers~\cite{amelio2020theory,zapletal2020,loirettepelous2021linearized}.
With respect to the laser linewidth, the method has already been used in
\cite{zhang2018phonon}  for two coupled resonators close to the exceptional point and in \cite{amelio2020theory} to compute the Petermann factor of a topological laser. 
Here a more general presentation of the method is given and 
it is shown to provide
a very convenient treatment
when spatial inhomogeneities and refractive index nonlinearities interplay.
One main insight  is that the Petermann and Henry factors can be explained in a unifying perspective as both arising from the  non-orthogonality of the Bogoliubov modes, which live in a space with doubled degrees of freedom.

This technology is then applied to the fluids of light~\cite{carusotto2013}, having particularly in mind polariton condensates~\cite{kasprzak2006bose}. While the Bogoliubov method is routinely applied to study the fluctuation modes in these systems \cite{wouters2007} and the linewidth is typically large enough to be easily measured~\cite{love2008,whittaker2009}, the Petermann broadening of the linewidth has never been considered in this context. In our numerical study
we consider different trapping regimes and pump spot sizes
to elucidate the role of gain guiding and interactions in shaping the lasing mode and determining the linewidth.
In particular we point out that, while Henry's mechanism (i.e. phase diffusion driven by intensity and refractive index fluctuations) catches most of the linewidth broadening, it is not sufficient to consider the density dynamics in the single mode approximation.
 
The paper begins with a review of the traditional theory of the linewidth, then we introduce our method in full generality and finally we apply it to polariton fluids.

\section{Linewidth theory}

We consider a general complex Ginzburg--Landau equation (CGLE) of the form
\begin{equation}
i\partial_t\psi(\vec{x},t) = \mathcal{W}[n] \psi(\vec{x},t) + \sqrt{D(\vec{x})} \xi
\label{eq:CGLE}
\end{equation}
where the operator $\mathcal{W}[n]$ can depend on position and contain differential operators, but  only depends on the field through its instantaneous density $n = |\psi|^2$.
Here we restrict to classical white noise satisfying 
$ 
\langle \xi^*(\vec{x},t) \xi(\vec{x}',t') \rangle =
2 \delta(\vec{x}-\vec{x}') \delta(t-t')
\ , \ \ \
\langle \xi(\vec{x},t) \xi(\vec{x}',t') \rangle = 0
.$

The steady-state lasing mode $\psi_0(\vec{x},t)$ satisfies $i\partial_t \psi_0(\vec{x},t) = \mathcal{W}[n_0] \psi_0(\vec{x},t)
= \omega_0 \psi_0(\vec{x},t)$ 
with 
$\omega_0$ the real-valued laser frequency.

\subsection{Henry's linewidth}

A naive derivation of the linewidth can be performed by allowing the intensity and phase of the lasing mode to fluctuate and neglecting other spatial modes. Two scalar equations for the phase  and density fluctuations are obtained integrating Eq.~(\ref{eq:CGLE}) over $\int d\vec{x} \ \psi^*_0(\vec{x})$, respectively
\begin{equation}
\partial_t \phi = \bar{\mu} I + \sqrt{\frac{\bar{D}}{n_{tot}}} \xi_{\phi}, \ \ \
\partial_t I = -\bar{\Gamma} I + 2\sqrt{\frac{\bar{D}}{n_{tot}}} \xi_{I}.
\label{eq:phase_density_dynamics}
\end{equation}
 Here
the bar stands for spatial averaging weighted with the steady-state density 
 $
\bar{f} \equiv \frac{1}{n_{tot}} \int d\vec{x} \ n_0(\vec{x}) f(\vec{x})
$
and the (averaged) interaction strength and density relaxation rate are
\begin{equation}
\bar{\mu} = \overline{ {\rm Re}\frac{\partial\mathcal{W}}{\partial n} n_0 } 
\ ,   \ \ \ \
\bar{\Gamma} = -2 \overline{   {\rm Im}\frac{\partial\mathcal{W}}{\partial n} n_0 }.
\label{eq:bars}
\end{equation}

Integrating Eqs. (\ref{eq:phase_density_dynamics}) yields the Henry linewidth \cite{henry1982}
\begin{equation}
\gamma_{H} = \frac{\bar{D}}{n_{tot}}(1 + \alpha^2)
\label{eq:gammaHenry}
\end{equation}
where
the first factor $\gamma_{ST} = \frac{\bar{D}}{n_{tot}}$
is the Schawlow-Townes result and
$\alpha = 2 \bar{\mu} / \bar{\Gamma}$ is the Henry factor correction. 
The linewidth is the diffusion rate of the phase, related to the field correlation function via 
$
g^{(1)}(t) \sim \exp\left[{-\frac{\gamma_{H}}{2}t}\right]
$.

\subsection{Petermann's factor}

The main problem with this approach is that projecting through $\int d\vec{x} \ \psi^*_0(\vec{x})$ is justified only for linear hermitian problems, where all the modes are mutually orthogonal. As first noted by Petermann~\cite{petermann1979}, the modes of a laser device are generally not orthogonal and spontaneous emission into these other  modes will eventually result in ``excess'' noise in the laser mode.
The total linewidth would  be given by the Schawlow-Townes result broadened by the product of the Henry and Petermann factor $\mathcal{K}_0$ 
\begin{equation}
\gamma_{H \times P} = \mathcal{K}_0 \times (1+\alpha^2) \times \gamma_{ST}
=
\mathcal{K}_0 \times \gamma_{H}.
\label{eq:gamma_HxP}
\end{equation}

The traditional treatment of the Petermann broadening requires the  strong assumption of being close to the lasing threshold, where the nonlinear device approaches a linear amplifier. This provides us the linear operator 
$\mathcal{W}[0]$ of which we can consider the right eigenmodes 
$\hat{\mathcal{W}} \varphi_k =  \omega_k \varphi_k$
 and their adjoints  (or left eigenmodes)
 $\hat{\mathcal{W}}^{\dagger} \tilde{\varphi}_k
  =  \omega_k^* \tilde{\varphi}_k$.

With the extra assumption that noise has uniform strength (which is a reasonable hypothesis in a uniformly pumped Fabry-Perot laser with inhomogeneous losses concentrated at the mirrors), one can 
perform the projection via 
 $\int dx \ \tilde{\varphi}_G^*$ 
 and
arrive at a concise expression for $\mathcal{K}_0$: 
 \begin{equation}
 \mathcal{K}_0 =
 \int d\vec{x} \ \tilde{\varphi}_0^* \tilde{\varphi}_0.
 \label{eq:K0}
 \end{equation}
Since we assumed the normalizations
 $
 \int d\vec{x} \ \tilde{\varphi}_k^* \varphi_l = \delta_{kl}
 $ 
  and $\int d\vec{x} \ {\varphi}_k^* \varphi_k = 1$, the adjoint modes are not simultaneously normalizable and in general $ \mathcal{K}_0 \geq 1$.
A Petermann factor of about 1.5 was first observed by Hamel and Woerdman
~\cite{hamel1990} in a semiconductor laser with large outcoupling.
 
In addition to the strong assumptions mentioned above, this treatment, described by
\cite{hamel1989,grangier1998,berry2003}, also requires to neglect the  refractive index nonlinearities.
On the other hand, the treatment by Henry \cite{henry1986}, which includes also $\alpha \neq 0$,  is still based on the Green function of the cavity in the absence of the laser field and close to threshold, so that the refractive index is not self-consistently reshaped by the presence of the laser field.

In the last decade, these shortcomings have stimulated   several works aimed at considering nonlinearities~\cite{chong2012}, multimode lasing \cite{pick2015}  and quantum effects \cite{pick2019} by means of a scattering matrix method. The review of this approach and the comparison with the treatment sketched below goes beyond the scope of the present paper.

\subsection{Bogoliubov method}

Here we illustrate an alternative and very concise approach based on the  Bogoliubov method and anticipated in \cite{amelio2020theory}. For a lattice of $N$ sites labeled as $\vec{x}$ in arbitrary dimensionality (for a continuous system one can consider a spatial grid, as typically done when solving numerically the CGLE.), let us call $\mathcal{L}_{las}$ the $2N\times 2N$ Bogoliubov matrix of the linearized dynamics on top of the lasing steady-state:
\begin{equation}
i\partial_t \begin{pmatrix}
\delta\psi \\ \delta\psi^*
\end{pmatrix} = \mathcal{L}
\begin{pmatrix}
\delta\psi \\ \delta\psi^*
\end{pmatrix} + \sqrt{D(\vec{x})}
\begin{pmatrix}
\xi \\ -\xi^*
\end{pmatrix},
\label{eq:BogoEq}
\end{equation}
where, if for notational  simplicity~\footnote{Since the phase freely diffuses one cannot just integrate the noise from $t = -\infty$ but a given phase must be arbitrarily chosen at an initial time. In practice, we can just take the initial state unperturbed because density fluctuations will decay exponentially fast; if one takes a perturbed initial state there will be some extra transient in Eq.~(\ref{eq:delta_psi}), which does not contribute to the linewidth. When considering only the density fluctuations, like in Eq.~(\ref{eq:delta_n}) below, one can integrate from $t = -\infty$ without need of an initial condition.} we set the unperturbed initial condition $\psi(\vec{x},0) = \psi_0(\vec{x})$, the field fluctuation is defined from 
$\psi(\vec{x},t) = (\psi_0(\vec{x}) + \delta\psi(\vec{x},t))e^{-i\omega_0t}$.

Let $V=\{  V_{\vec{x}\sigma,a} \}$ be the invertible matrix which diagonalizes $\mathcal{L}_{las}$, where {the pseudo-spin} $\sigma =\uparrow,\downarrow$  {indicates} the particle and hole components of the Bogoliubov problem and $a$ labels the $2N$ eigenmodes.
The Goldstone mode $V_{\vec{x}\sigma,G}$, that we assume to be unique with all other excitations having a finite life-time, is the eigenstate with zero eigenvalue. As usual, its spatial shape follows the one of the lasing mode  and we can 
set 
$(V_{\vec{x}\uparrow,G},V_{\vec{x}\downarrow,G}) 
= \frac{1}{\sqrt{2n_{tot}}}
(\psi_0(\vec{x}),  -\psi_0^*(\vec{x}))$ as a natural gauge choice.
Notice that each Bogoliubov eigenmode, or each column of $V$, is defined modulo a phase factor, hence the use of ``gauge'' word.

The Bogoliubov problem  (\ref{eq:BogoEq}) reduces to $2N$ independent equations upon multiplication by $V^{-1}$; it is then straightforward to integrate  the field
\begin{equation}
\delta \psi(\vec{x},t) 
= \sum_{a } \int_0^t ds \ 
\pi_a(s) e^{-i\omega_a (t-s)},
\label{eq:delta_psi}
\end{equation}
where  $\omega_a$ are the complex Bogoliubov eigenvalues.
The overlap  $\pi_a = -i\sum_{\vec{x}\sigma} V^{-1}_{a,\vec{x}\sigma} \sqrt{D(\vec{x})} \xi_{\vec{x}\sigma}$, with $\xi_{\vec{x}\sigma} = (\xi_{\vec{x}}, - \xi_{\vec{x}}^*)$, represents the projection of noise on a given mode.
We also define $\pi_{ab}$ from
$\langle
\pi_a(t) \pi_b(t') \rangle =
2 \delta(t-t')
\sum_{\vec{x}} D(\vec{x})
\left(
V^{-1}_{a, \vec{x}\uparrow} V^{-1}_{b\vec{x}\downarrow} + 
V^{-1}_{a, \vec{x}\downarrow} V^{-1}_{b\vec{x}\uparrow}
\right)
\equiv \pi_{ab} \delta(t-t')
$.
One can switch to a density-phase formalism: the density and phase contributions of the $a$-th  mode are respectively
\begin{equation}
  \nu_a(\vec{x}) =  \psi_0^*(\vec{x}) V_{\vec{x}\uparrow,a} +
\psi_0(\vec{x}) V_{\vec{x}\downarrow,a} 
\end{equation}
and 
\begin{equation}
  \phi_a(\vec{x}) =  \frac{i}{2n(\vec{x})}
  \left[
  \psi_0^*(\vec{x}) V_{\vec{x}\uparrow,a} -
\psi_0(\vec{x}) V_{\vec{x}\downarrow,a}
\right]. 
\label{eq:Bogo_phase_component}
\end{equation}
The particle-hole symmetry of the Bogoliubov matrix~\cite{amelio2020galilean} guarantees that $\nu_a, \phi_a$ are real-valued.
For the Goldstone mode $a=G$ 
one has that
$\nu_G=0$ and  $\phi_G(\vec{x}) \equiv \phi_G = -i/\sqrt{{2 n_{tot}}} $ is a constant, consistently with the fact that the Goldstone mode corresponds to a global phase rotation.

The crucial observation is that, since all modes $a \neq G$
decay exponentially in time, in the large $t$ limit
the phase-phase correlator gets a constant contribution from the $a \neq G$ modes plus
the linear in $t$ contribution 

\begin{equation}
\langle
\delta\phi^2
\rangle
\sim
\pi_{GG} \phi_G^2 t
\label{eq:dphi2BOGO}
\end{equation}
coming from the Goldstone mode, hence
\begin{equation}
\gamma_{Bogo} = \pi_{GG} \phi_G^2 =  
- \frac{2}{n_{tot}}
 \sum_{\vec{x}} D_{\vec{x}} \ (V^{-1})_{G,\vec{x}\uparrow} (V^{-1})_{G,\vec{x}\downarrow} .
\label{eq:gammaBOGO}
\end{equation}
Notice that the very last expression is not gauge invariant, but  assumes the Goldstone mode to be defined as $(V_{\vec{x}\uparrow,G},V_{\vec{x}\downarrow,G}) 
= \frac{1}{\sqrt{2n_{tot}}}
(\psi_0(\vec{x}),  -\psi_0^*(\vec{x}))$;
in particular, in this gauge the minus sign gets cancelled by the product  $(V^{-1})_{G,\vec{x}\uparrow} (V^{-1})_{G,\vec{x}\downarrow}$.
Equations (\ref{eq:Bogo_phase_component}) and (\ref{eq:dphi2BOGO})  are instead in a gauge covariant and invariant form, respectively.

Notice that $\mathcal{L}$ is typically a very sparse matrix and  $(V^{-1})^{\dagger}$ are the eigenmodes of 
$\mathcal{L}^{\dagger}$.
For very large numerical grids, the Goldstone column, corresponding to the zero eigenvalue,
can be efficiently computed via the Lanczos algorithm.
However, the Lanczos finds the normalized column 
$[(V^{-1})^{\dagger}]_{\vec{x}\sigma,G}$,
while 
$(V^{-1})_{G,\vec{x}\sigma}$ is in general not normalized if 
$V_{\vec{x}\sigma,G}$ is. The correct normalization, which  is fundamental here, can be found by requiring 
$\sum_{\vec{x},\sigma}
 (V^{-1})^{\dagger}_{\vec{x}\sigma,G} V^*_{\vec{x}\sigma,G} = 1$  with 
 $\sum_{\vec{x},\sigma}
|V_{\vec{x}\sigma,G}|^2 = 1$ 
 .

\begin{figure*}[t]
\centering
\includegraphics[width=0.44\textwidth]{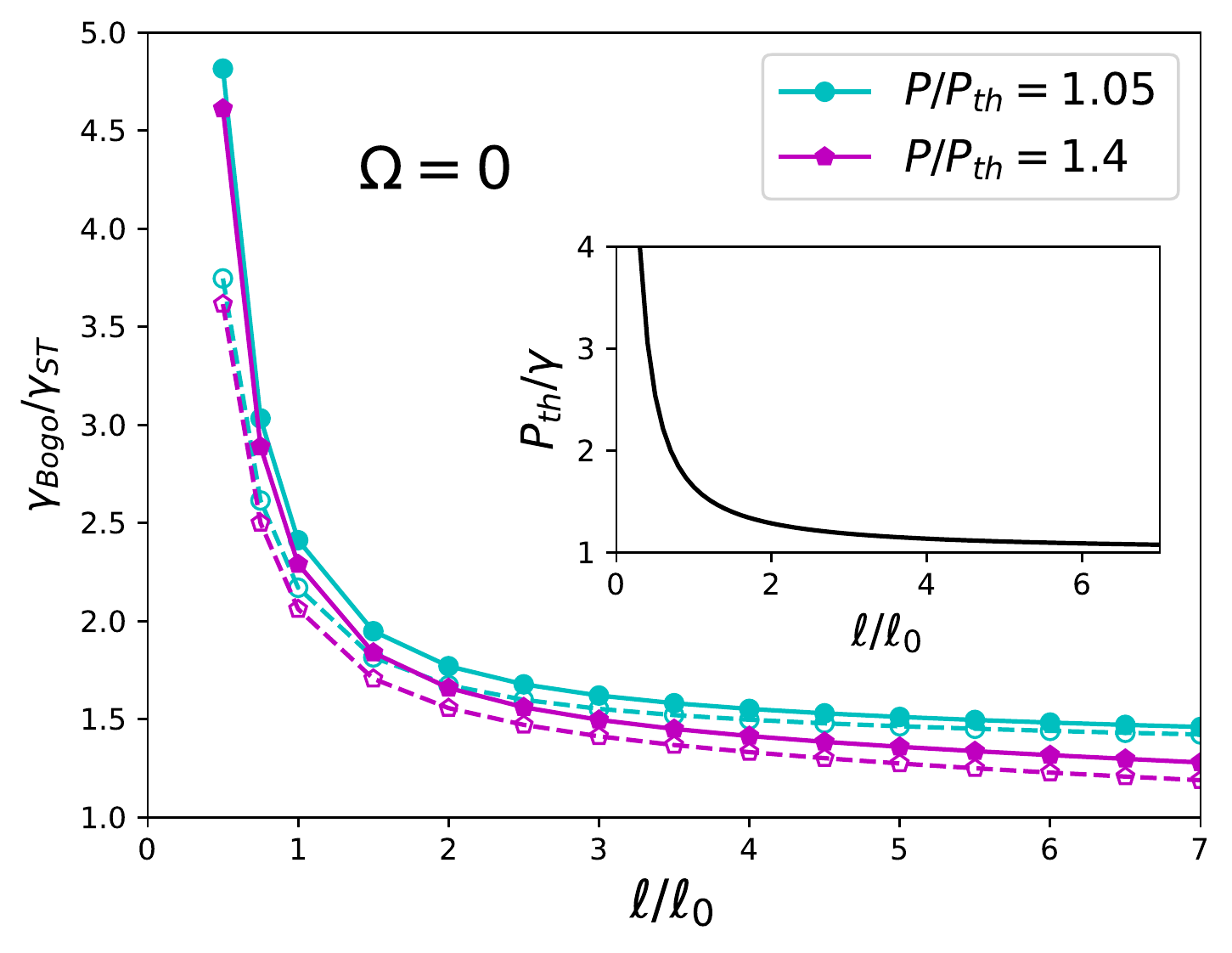}  
\includegraphics[width=0.45\textwidth]{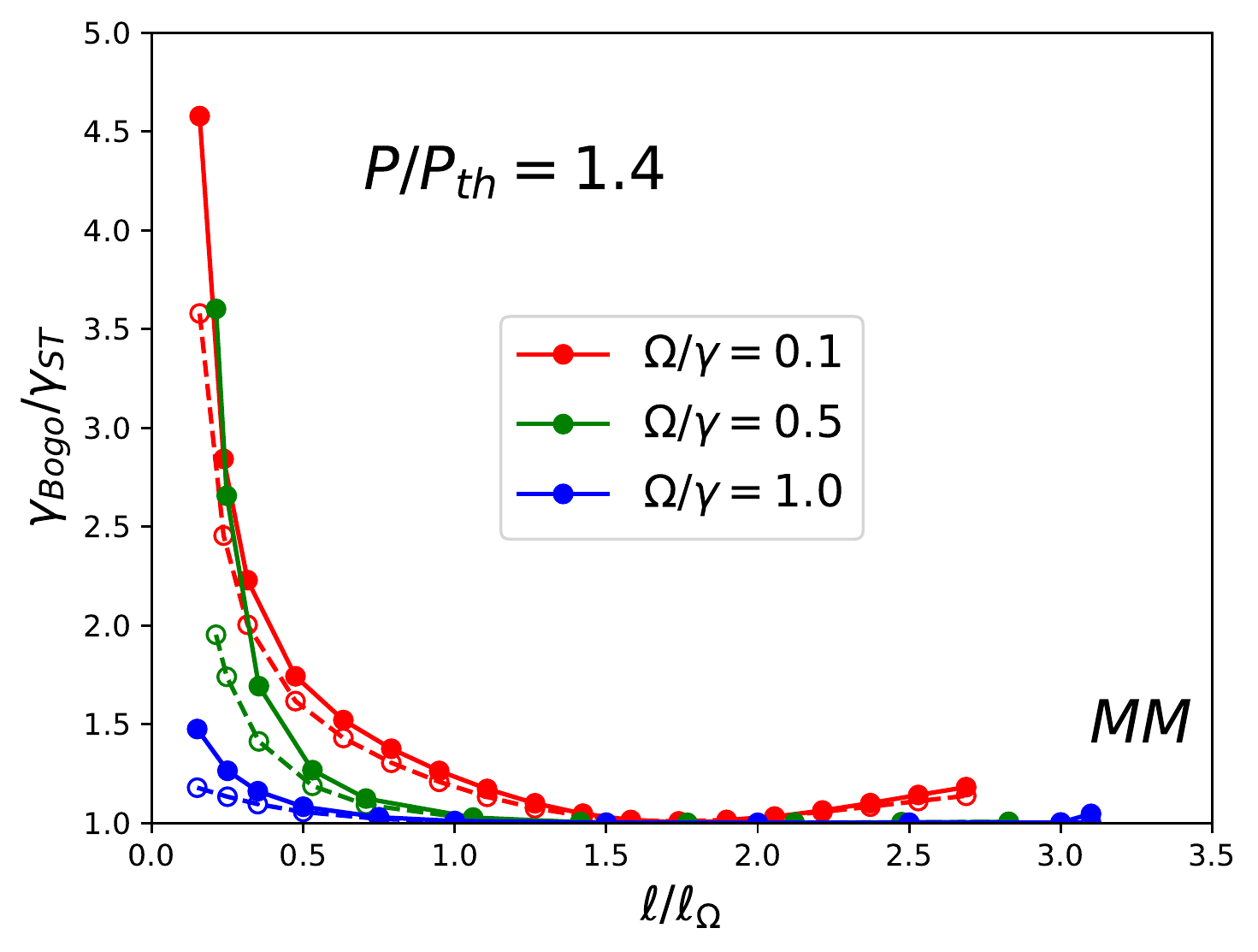}
\caption{  Petermann-like broadening  $\gamma_{Bogo}/\gamma_{ST}$   
as a function of the pump-spot width $\ell$  for the model of Eq.~(\ref{eq:GPE}) with $g=0$.
The full circles correspond to $\gamma_{Bogo}/\gamma_{ST}$ computed according to Eqs. (\ref{eq:gammaBOGO}) and (\ref{eq:ST_VV1}), while the empty circles denote $\mathcal{K}_0$ obtained from $\mathcal{W}[n_0]$ according to  formula (\ref{eq:K0}), which assumes uniform noise and closeness to threshold.
(left) The case without trapping $\Omega=0$ is shown as a function of $\ell/\ell_0$ and for two different pump powers (normalized to threshold). For small pump spots the strong gain guiding introduces a lot of excess noise. Inset: the pumping threshold also increases for smaller spots, since a larger outflow has to be compensated.
(right) The harmonic trap yields another length scale $\ell_{\Omega}$. For stronger trapping, the  modes of the system resemble the eigenstates of the oscillator Hamiltonian and excess noise is reduced. 
 For $\ell/\ell_{\Omega} \sim 3$ higher harmonics will lase, leading to a multimode (MM) phase with no steady-state.
}
\label{fig:gain_guiding}
\end{figure*}

\subsection{Discussion}

Crucially, if the adjoint mode to the Goldstone is the Goldstone itself  $(V^{-1})^*_{G,\vec{x}\sigma} = V_{\vec{x}\sigma,G}$ (a sufficient condition for this is unitarity $VV^{\dagger} = 1$), the simple Schawlow-Townes expression holds for the linewidth\cite{schawlow1958}:
\begin{equation}
\gamma_{ST} = \frac{\bar{D}}{n_{tot}}
\label{eq:ST_VV1}
\end{equation}
with $n_{tot} = \sum_{\vec{x}}  n_{\vec{x}}$ and  $\bar{D} = \sum_{\vec{x}} D_{\vec{x}} n_{\vec{x}} / \sum_{\vec{x}}  n_{\vec{x}}$.

In  the case of a spatially uniform system lasing in the $\vec{k}=0$ mode with amplitude $\sqrt{n_0}$,   the different wavevectors decouple in the Bogoliubov problem
and the $2\times 2$ sector  $\mathcal{L}_{las}(\vec{k}=0)$ is diagonalized by the   matrix
\begin{equation}
V_{\sigma,p}(\vec{k}=0) = \frac{1}{\sqrt{2}}
\begin{pmatrix}
 1& \frac{1 - i\alpha}{\sqrt{1+\alpha^2}} \\
-1 & \frac{1 + i\alpha}{\sqrt{1+\alpha^2}} \\
\end{pmatrix}
\label{eq:HenryVmatrix}
\end{equation}
(the first column  $p=G$  here is the Goldstone mode
and 
$[V^{-1}(k=0)]_{G,\sigma}=
\frac{1}{\sqrt{2}}
(1+i\alpha, -1+i\alpha)
$
)
from which one recovers the Henry formula Eq.~\ref{eq:gammaHenry}.
In particular, notice that for  $\alpha=0$ the mode matrix $V$ is unitary: the Henry factor can then be interpreted as the non-orthogonality  of the modes of (\ref{eq:HenryVmatrix})  with respect to the particle-hole index (as opposed to the spatial or momentum index).
The same conclusions apply to a point-like laser, which physically means that the system dynamics  is essentially described by using a single mode of a lossless cold cavity. Alternatively, one can reason in terms of many conservative modes that are well separated in energy, while losses and pumping couple different modes. If the coupling is negligible one can stick to a single mode approximation. If also $\alpha=0$,  the sector corresponding to the lasing mode features  $(V^{-1})^*_{G,\vec{x}\sigma} = V_{\vec{x}\sigma,G}$ and there is no broadening.

In the opposite scenario, the  refractive index nonlinearity ${\rm Re} \frac{\partial \mathcal{W}}{\partial n}$ is negligible, but gain and losses strongly couple different conservative modes of the idealized cavity.
This is the generic case for a  system with localized gain or losses and closely spaced conservative modes (i.e a large cavity or a weak trapping potential).
In this case,  the broadening of the linewidth with respect to Eq.~(\ref{eq:ST_VV1}) will be interpreted as a Petermann-like factor, since the non-unitarity of $V$ comes from the coupling of different conservative modes.
However, we stress here that in general the two broadening mechanisms will  not  be separable.
This will be particularly true where the refractive index and gain profiles are modified by the intense field with respect to the cold cavity, so that both the real and imaginary part of $\mathcal{W}$ contribute nonlinearly to the shape of the lasing mode.

As a final remark, it is worth reminding that the Bogoliubov theory presented here is only valid in the high spatial coherence regime, where 
KPZ-like effects and the broadening related to the lack of long-range order are negligible~\cite{amelio2020theory}.

\section{Polariton condensates}

As an illustrative example of the general theory,
we now move to consider the linewidth of a condensate of light~\cite{carusotto2013}, with special focus on the case of incoherently driven polariton fluids~\cite{kasprzak2006bose}. 
For simplicity we will restrict our analysis to the so called adiabatic scenario where the dynamics of carriers is traced out, so that
one is left with the following CGLE~\cite{wouters2007}:
\begin{multline}
i\partial_t \psi = \left[ 
-\frac{1}{2m}\nabla^2 + g n + V(x) + \right. \\
\left. + \frac{i}{2}
\left( 
\frac{P(x)}{1+n/n_S} - \gamma (x)
\right)
\right] \psi +  \sqrt{D(x)} \xi,
\label{eq:GPE}
\end{multline}
which is linearized to the Bogoliubov matrix:

\begin{widetext}
\begin{equation}
\mathcal{L} =
\begin{pmatrix}
-\frac{1}{2m}\nabla^2 + 2 g n_0 + V -\omega_0 + \frac{i}{2}
\left( 
\frac{P}{1+n/n_S} - \gamma 
\right)
 -\frac{i}{2} \Gamma 
&
g \psi_0^2  - \frac{i}{2} \tilde{\Gamma}
\\
-g (\psi_0^*)^2  - \frac{i}{2} \tilde{\Gamma}^* & 
+\frac{1}{2m}\nabla^2 - 2 g n_0 - V +\omega_0 + \frac{i}{2}
\left( 
\frac{P}{1+n/n_S} - \gamma 
\right) -\frac{i}{2} \Gamma 
\end{pmatrix}
\end{equation}
\end{widetext}


Here $\Gamma(x) = \frac{ n_0(x)/n_S}{(1+n_0(x)/n_S)^2} P(x)$ gives
  the  local rate for density relaxation  and depends  on the density itself;
  we also used
  $\tilde{\Gamma}(x) = \frac{ \psi_0^2(x)/n_S}{(1+n_0(x)/n_S)^2} P(x)$.
  In the following we will assume  homogeneous losses $\gamma$ and a Gaussian pump spot $P(x)=P\exp(-x^2/2\ell^2)$ of width $\ell$.
 Noise will be taken with a diffusion shape $D(x) \propto P(x)$, in analogy with spontaneous emission.
 For the external potential we take a harmonic trap $V(x)=\frac{m}{2} \Omega^2 x^2$.
 
 While here
 for simplicity
 we will report  results in 1D, we expect that the physics remains qualitatively the same in 2D.

\subsection{Non-interacting untrapped condensate}
 
As a first step, we will neglect the refractive index nonlinearity and set $g=0$.
In the absence of trapping ($\Omega=0$), there are two typical scales that compete: the width of the pump $\ell$ and the wavelength $\ell_0 =  1/\sqrt{m\gamma}$, corresponding to a kinetic energy of the order of the dissipation rate. 
Lasing is possible when the fluid can be localized by the pump spot, a situation we will refer to as {\em gain guiding}.
Small $\ell/\ell_0$ means that the pump is trying to excite lasing in a narrow region: the large outflowing current makes the effective loss rate of the  lasing mode much greater than $\gamma$. Correspondingly, $P_{th}/\gamma \gg 1$, as shown in the inset of Fig. \ref{fig:gain_guiding}.a. For  $\ell \to \infty$ the power threshold approaches $\gamma$, the result for a uniform system where the flow of the fluid is negligible.

Moreover, generating a fluid in a region small compared with $\ell_0$ comes with a price in terms of coherence. This is shown in Fig. \ref{fig:gain_guiding}.a, where the ratio $\gamma_{Bogo}/\gamma_{ST}$ is plotted
in solid dots
as a function of $\ell/\ell_0$  for $P/P_{th}=$1.05 (cyan) and 1.4 (magenta); as expected, it diverges for small $\ell/\ell_0$  and goes to 1 for large spots. In this setting, $\gamma_{Bogo}/\gamma_{ST}$ can be safely interpreted as a Petermann factor.
The empty circles refer to $\mathcal{K}_0$ computed from $\hat{\mathcal{W}}[n_0]$ according to Eq.~(\ref{eq:K0}), which provide a good approximation in this regime.
As it is apparent from the plot, the results depend weakly on $P/P_{th}$ (actually $P$ enters in $\bar{\Gamma}$ and provides a third length scale  $\ell_{P} \sim  1/\sqrt{m\bar{\Gamma}}$; zooming in on the tails of Fig.~\ref{fig:gain_guiding} one could see that $\gamma_{Bogo}/\gamma_{ST}$ gets very close to 1 for $\ell \gg \ell_{P}$  (not shown)). 

\subsection{Non-interacting trapped condensate}

For $\Omega>0$, also the length scale 
$\ell_{\Omega} =  1/\sqrt{m\Omega}$ has to be taken into account, yielding the interesting behavior of $\gamma_{Bogo}/\gamma_{ST}$ shown in Fig. \ref{fig:gain_guiding}.b.
For  $\ell,\ell_0>\ell_{\Omega}$ trapping is very effective and the lasing mode 
 approaches the ground-state of the harmonic oscillator: this is the regime of conservative cavity confinement and the Petermann-like factor is close to 1. The pump term couples higher harmonic modes only perturbatively, so that at zeroth order the modes of the systems are orthogonal. On the contrary, for weak trapping (large $\ell_{\Omega}/\ell_0$) and narrow pump spots (small $\ell/\ell_{\Omega}$ i.e. small $\Omega/\gamma$) we recover the regime of gain guiding described above for $\Omega=0$ and the corresponding divergence of the linewidth. We can think of this regime as the very localized gain deforming the harmonic ground state; on the other hand, when 
 $\ell/\ell_{\Omega} \sim 1.75$
 the pump spot and the ground state of the harmonic trap fit each other in size and 
 $\gamma_{Bogo}/\gamma_{ST}$
 is very close to 1. Anyway, even for small $\ell/\ell_{\Omega}$ the conservative limit is eventually recovered for $\Omega \gg P$.

An increase of the Petermann-like factor occurs approaching $\ell/\ell_{\Omega} \sim 3$ and an instability develops for larger pump spots (for which the data are truncated).
Indeed, if the pump diameter $\ell$ is larger than the harmonic oscillator wave-function, two or more conservative modes can feel enough gain and be enough spatially separated to lase simultaneously.
As a result, there is no steady-state and the emission spectrum consists of different peaks.
Related physics was reported in 2D polariton systems, where a steady-state can be reached in which  the rotational symmetry is spontaneously broken by the formation of an array of vortices~\cite{keeling2008}.
The coherence properties of this
multi-mode regime go beyond the scope of the present work.

\subsection{Interacting trapped condensate}

\begin{figure}[t]
\centering
\includegraphics[width=0.47\textwidth]{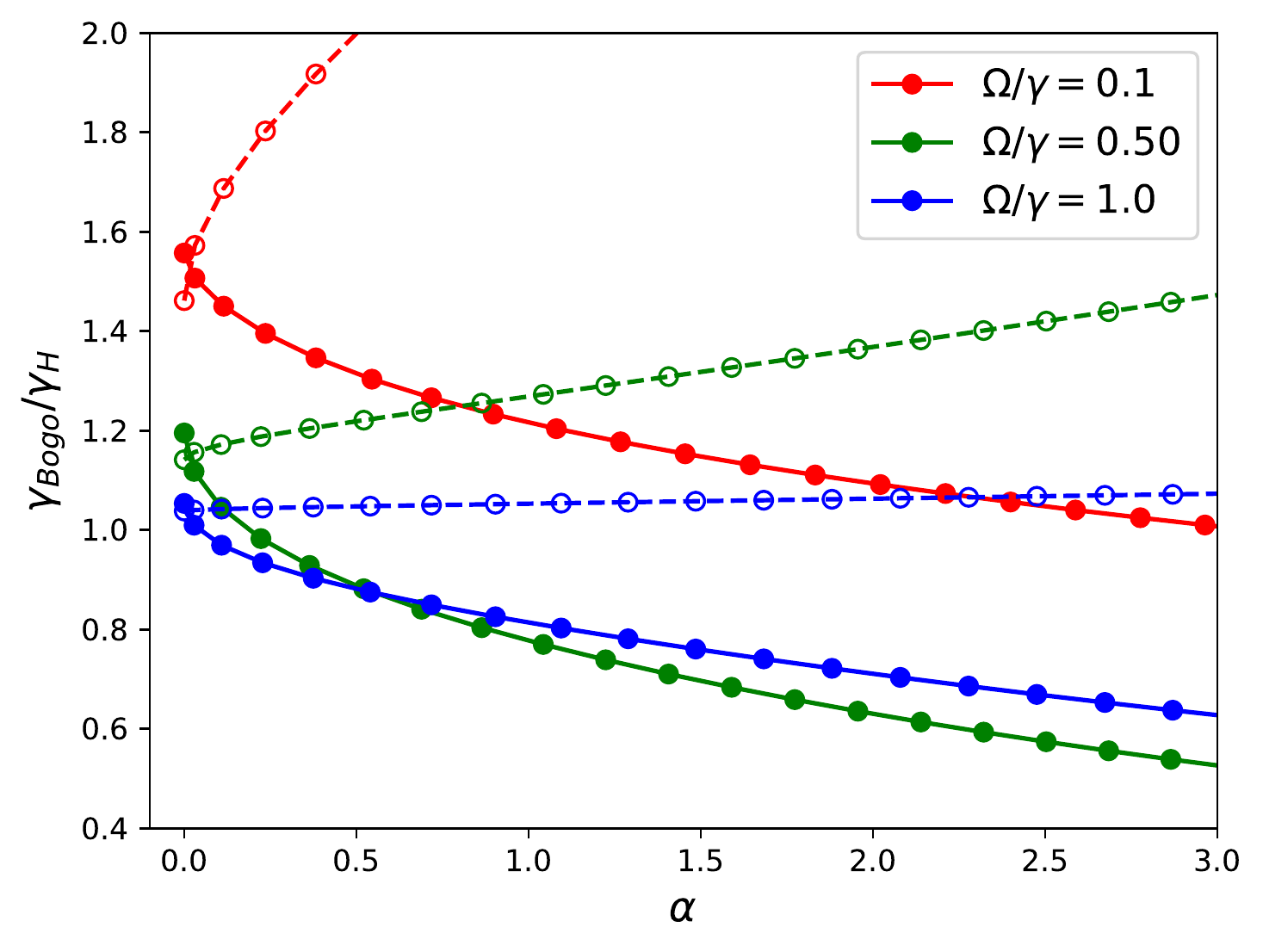}  
\caption{ Ratio $\gamma_{Bogo}/\gamma_{H}$ (solid line and full circles) versus $\alpha$, which quantifies the relevance of refractive index fluctuations and corresponds to increasing the nonlinearity $g$. Different trapping frequencies $\Omega$ are reported; for larger $\Omega$, one can see that 
the single mode approximation holds for small enough $\alpha$, so that $\gamma_{Bogo}/\gamma_{H\times P}$ gets close to one, but this argument fails at finite $\alpha$, signalling a limitation of the Henry's approach.
Empty symbols and dashed lines correspond to $\mathcal{K}_0$ from the standard treatment of Eq.~(\ref{eq:K0}): it is evident that  the
standard Petermann times Henry factor treatment of Eq.~(\ref{eq:gamma_HxP}) does not work at all in the presence of interactions, calling for the full Bogoliubov treatment. 
Here we have fixed $P=2\gamma$ and pump diameter $\ell/\ell_0 \simeq 2.8$.
}
\label{fig:interactions}
\end{figure}

When $g>0$ the Henry factor can give a very large contribution to the linewidth. In Fig. \ref{fig:interactions} we plot 
 $\gamma_{Bogo}/\gamma_{H}$ 
versus $\alpha$ for different trapping strengths. Each data set is obtained by increasing the refractive index nonlinearity $g$, which results in the Henry factor 
$\alpha=2\bar{\mu}/2\bar{\Gamma}$,
computed inserting into Eq.~(\ref{eq:bars}) the  steady-state profile found numerically for each $g$.

Trapping,  gain and interactions contribute in a complex way to the shape of the laser mode and, as a consequence, to the linewidth. 
The first consideration is that
the Henry prediction $\gamma_H$ is at order zero a decent approximation to the linewidth  (we recall that the linewidth broadening goes like $1+\alpha^2$ and can be very large, so that for $\alpha \sim 3$ one expects a broadening by a factor of $\sim 10$ and the error of the green and blue data by a factor 2 is still better than using the vanilla Schawlow-Townes linewidth). 
However, for a given $\alpha$, the ratio $\gamma_{Bogo}/\gamma_{H}$ does not get monotonically closer to 1 when increasing the trapping $\Omega$, as it would be expected in virtue of an improved single-mode approximation.
The issue with this poor performance of the $\gamma_H$ approximation
will be elucidated in the next subsection.

Before that, we discuss the behavior of the traditional Petermann factor
$\mathcal{K}_0$ from Eq.~(\ref{eq:K0}).
Looking at the fast blowing up of the red empty circles,  it 
 is apparent that the
standard expression $\gamma_{H \times P}$ in Eq.~(\ref{eq:gamma_HxP}) completely misses the exact result,  overestimating the linewidth, more and  more dramatically  for larger $g$.
Therefore, Eq.~(\ref{eq:K0}) seems to be useless for the interacting system.

\subsection{Improved Henry's scheme}

\begin{figure}[t]
\centering
\includegraphics[width=0.47\textwidth]{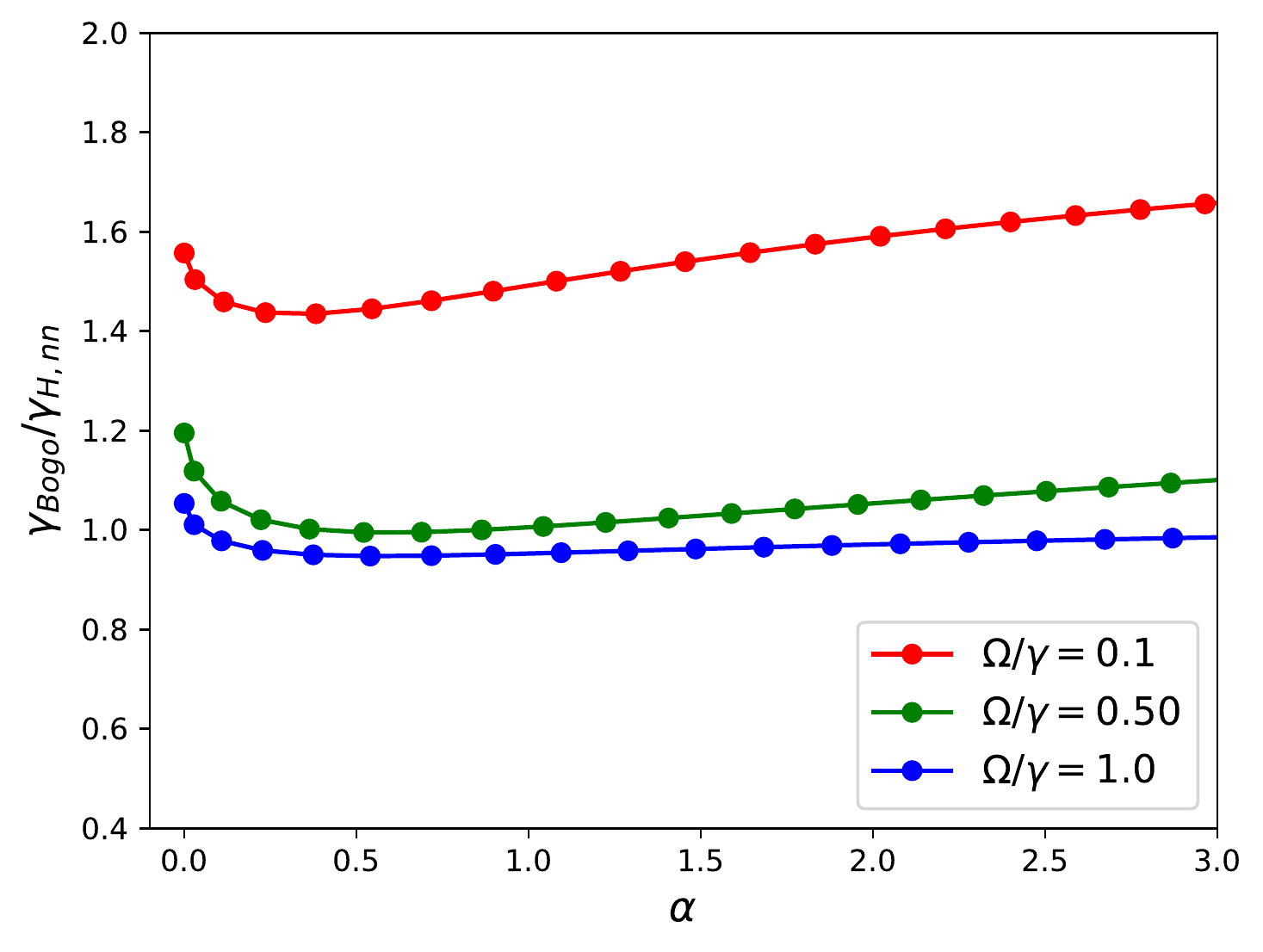}  
\caption{
Ratio $\gamma_{Bogo}/\gamma_{H,nn}$ for increasing interactions.
The regime is exactly the same of Fig.~\ref{fig:interactions}, but now $\gamma_{H,nn}$ accounts for the density driven drift of the phase  with the density fluctuations computed beyond the single mode approximation.
This shows that Henry's intuition is physically correct, but that density dynamics is microscopically complex.
Importantly,  $\gamma_{Bogo}/\gamma_{H,nn}$ goes to 1 for increasing trapping.
}
\label{fig:nn}
\end{figure}

As we show below, the  quantitative disagreement of $\gamma_H$ and $\gamma_{Bogo}$ highlighted in Fig.~\ref{fig:interactions} should not undermine
Henry's intuition that density fluctuations have the greatest impact on the drift of the phase. 
The issue with Fig.~\ref{fig:interactions}
is that $\gamma_H$ in the Henry's scheme is computed in the single mode approximation, while density fluctuations are notoriously short ranged and short lived~\cite{chiocchetta2013}. Here we show that Henry's mechanism accounts quantitatively for the exact linewidth provided the  phase mode is driven by the full density dynamics.
This means that the density-density correlator should be computed retaining all the Bogoliubov modes.

For this purpose, we make use of the ansatz
\begin{equation}
\psi(x,t) = \psi_0(x,t) \left[ 
1 + \frac{\delta n(x,t)}{2 n_0(x)}  - i\phi(t) + ... \right].
\label{eq:ansatz}
\end{equation}
Crucially only the global phase fluctuations are considered, while the full spatial dependence of the density dynamics is taken into account.
The spatial uniformity of $\phi(t)$ is  the only approximation we  make in this subsection and which cuts off Petermann-like correlations.

After integration by 
$\int dx \ \psi^*_0(x)$ the equation for the phase becomes
\begin{equation}
\partial_t \phi = g \overline{\delta n}  + \sqrt{\frac{\bar{D}}{n_{tot}}} \xi_{\phi},
\label{eq:phase_pro}
\end{equation}
where we remind that according to the bar notation established above  $\overline{\delta n}$
is the spatially averaged density fluctuation with weight $n_0(x)$.
Once again we remark that this expression is not exact, but some non-orthogonality between the modes has been neglected.
However, it allows for a more refined expression for the linewidth with respect to Eqs.~(\ref{eq:phase_density_dynamics}), 
since it doesn't assume that density fluctuates in a perfectly correlated way across the system. 
After  integration, one gets at the leading order in $t$
\begin{equation}
\langle
(\phi(t) - \phi(0))^2
\rangle \simeq \frac{\bar{D}}{n_{tot}} t +
2 g^2 t  \int_0^{\infty} d\tau \
\langle
\overline{\delta n}(\tau) \  \overline{\delta n}(0)
\rangle .
\end{equation} 
For an inhomogeneous system the dynamics of $\overline{\delta n}$ is rather complex, since it contains contributions from all the modes.
Nonetheless, the density-density correlator
can be computed exactly using the full Bogoliubov theory or numerically via relatively cheap simulations of the duration of a few $\Gamma^{-1}$.


We now sketch the computation of
$\langle
\overline{\delta n}(\tau) \  \overline{\delta n}(0)
\rangle$
using Bogoliubov's method. Since the undamped Goldstone mode does not contribute to the density fluctuations, one has
\begin{equation}
\delta n (x,t) = \sum_{a \neq G} \int_{-\infty}^t ds \ 
\nu_a(x)
\pi_a(s) e^{-i\omega_a (t-s)},
\label{eq:delta_n}
\end{equation}
where the integral can start from $-\infty$ since the initial condition for $\delta n$ becomes irrelevant after a time of order  $\Gamma^{-1}$.
Upon integration, the correlator reads
\begin{equation}
\langle
\overline{\delta n (t)} \  \overline{\delta n(t')}
\rangle = -i
\sum_{a,b \neq G} \frac{\bar{\nu}_a \bar{\nu}_b \pi_{ab}}{\omega_a + \omega_b} 
e^{-i \omega_a |t-t'|} .
\end{equation}
Finally, the linewidth reads
\begin{equation}
\gamma_{H,nn} =
\frac{\bar{D}}{n_{tot}} +
g^2 \sum_{a,b \neq G} \frac{\bar{\nu}_a \bar{\nu}_b \pi_{ab}}{-\omega_a \omega_b} .
\end{equation}

In the case of a point-like or uniform laser there are only two $k=0$ modes,  denoted as $a=G,A$ ($A$ stands for ``amplitude''), and one has 
$\langle \overline{\delta n}(\tau) \  \overline{\delta n}(0)
\rangle 
= 2 \frac{ n_{tot} \bar{D}}{ \Gamma} e^{-\Gamma |\tau|},
\pi_{AA} = 2D(1+\alpha^2), \bar{\nu}_A =\sqrt{\frac{2n_0}{1+\alpha^2}}
 $ and $ \omega_A = -i\Gamma
$, which leads to the standard Henry linewidth.

~

The ratio $\gamma_{Bogo}/\gamma_{H,nn}$
is plotted in Fig.~\ref{fig:nn} for increasing $\alpha$ for exactly the same regime of Fig.~\ref{fig:interactions} (for comparison $\alpha$ is also defined in the same way, even though it is a definition motivated by the single-mode approximation).
From the approximate flatness of the curves, it is clear that for any $g$ the linewidth is well captured by the Henry mechanism  but for a factor, which, as it is more evident from the red points, is of the order of $\gamma_{Bogo}/\gamma_{H,nn}(g=0)$.   
Therefore,  one would be tempted to interpret the discrepancy between $\gamma_{Bogo}$ and $\gamma_{H,nn}$ as a Petermann-like effect.
However, notice that $\gamma_{Bogo}/\gamma_{H,nn}$ can be less than one (blue points), which is impossible in the standard theory of the Petermann factor.
 This supports our claim that in general there is no natural way of disentangling the non-orthogonality of the Bogoliubov modes due to spatial and particle-hole indices.

We have then shown that density fluctuations, if properly calculated beyond the single-mode approximation, account for most of the linewidth broadening.
Crucially, notice that for any $\alpha$ the ratio $\gamma_{Bogo}/\gamma_{H,nn}$ tends systematically to 1 for stronger trapping, in contrast to Fig.~\ref{fig:interactions}.

\section{Conclusions}

To summarize, we have presented a general linear response treatment of the linewidth of laser devices operating in a single mode regime with good spatial coherence. 
Our Bogoliubov scheme allows for a unified treatment of the Petermann and of the Henry broadening mechanisms in terms of the non-orthogonality of  modes which live in a space with doubled degrees of freedom.

As a most relevant example, we have applied this technique to calculate the linewidth broadening due to gain guiding and density fluctuations  for a polariton condensate described by a generalized Gross-Pitaevskii equations.
The complex interplay of the harmonic trapping, the pump spot size and the interactions has been discussed in detail.
In particular, we have shown that one needs to include density-density correlations beyond the single mode approximation in order to quantitatively account for the interaction-induced Henry-like broadening.

Since the linewidth of a polariton condensate is a easily measurable quantity~\cite{love2008,whittaker2009}, the main difficulty to experimentally probe the different broadening mechanisms appears to be the calibration of the experiment; indeed, it is not straightforward to measure  the number of polaritons in the cavity and the noise strength, so any physical conclusion should rely on ratios of linewidths in different regimes.

Finally, the Bogoliubov method presented here can be straightforwardly generalized to deal with more realistic microscopic models, accounting for the dynamics of the carriers by  rate equations \cite{wouters2007}
or by the full Maxwell-Bloch equations~\cite{zhang2018phonon}.
In future works, it would be also interesting to study quantum effects to the linewidth \cite{pick2019} within a Wigner representation formalism~\cite{berg2009}.



\section*{Acknowledgements}
We are grateful to Stefan Rotter for a few tips on the Petermann factor.
I. C.  acknowledges financial support 
from \iac{the H2020-FETFLAG-2018-2020 project "PhoQuS" (n.820392),} and 
from the Provincia Autonoma di Trento.

\bibliography{bibliography}

\end{document}